Report on the Workshop

# Joint Observatories Kavli Science Forum

held at ESO Vitacura, Santiago, Chile, 25–29 April 2022

Pascale Hibon[1]
Jesús Corral-Santana[1]
Itziar de Gregorio-Monsalvo[1]
Leopoldo Infante[2]
Elizabeth Humphreys[3]
John Blakeslee[4]

[1] ESO
[2] Las Campanas Observatory, Chile
[3] Joint ALMA Observatory, Chile
[4] NOIRLab, USA

The Joint Observatories Kavli Science Forum in Chile was organised in hybrid mode with the aim of encouraging collaborations, not only with the Chilean institutions, but also between the different observing facilities based in Chile. The meeting featured scientific talks showing results obtained with the astronomical facilities based in Chile, but significant time was also dedicated to round-table discussions on Life Balance, Diversity-Equity-Inclusion, and the Road Ahead (i.e., the future of those Chile-based facilities).

Chile-based observatories have been leading scientific research in several astronomical areas. The forum was organised around the highest-impact scientific results provided by those facilities over the last few years. The aim was to show how these different observatories have contributed to those advances in astrophysics and, with that goal in mind, the organising committee placed particular emphasis on the scientific involvement of the astronomers working at those observatories to achieve cutting-edge results. The intention was to organise a meeting to gather together both observatory staff (astronomers, scientists, fellows, students, engineers, operators etc.) and Chilean institute researchers (postdocs, students etc.) to present their research. This would also reinforce the scientific collaboration between observatories and Chilean research institutes, to examine common experiences and concerns, and to discuss different points of view on how to cope with similar challenges.

Each half-day of the meeting was dedicated to a specific field of astronomy that had seen major advances thanks to the observing facilities based in Chile, followed by dedicated time for discussion.

After an introductory speech that included a report from the Director of the Chilean Astronomical Society (SOCHIAS), Monday was dedicated to the transient sky with talks focusing on cataclysmic variables, multi-messenger follow-up, and supernovae. It was reported that new observing systems are being developed, focused on the management of time-domain observations. We would like to highlight the sharing of the different points of view, concerns and strategies between several observatories during this session. Representatives of all the observatories agreed that although there will be an enormous number of alerts/triggers at the beginning of the Legacy Survey of Space and Time, the rate of triggering will slow down and the tools will adapt, as will the observatories and the community.

Tuesday morning's talks focused on exoplanets and star formation. A great example was presented of successful ALMA+VLT science and how a Chilean-led research team is making great use of all the Chile-based facilities. During the afternoon we had our first round-table discussion, on Life Balance, chaired by Itziar de Gregorio-Monsalvo and with several invited panellists (see de Gregorio-Monsalvo, Hibon & Alcalde Pampliega on p. 44 of this issue for more details). The three key words in the discussion were flexibility, tolerance and empathy, and the conclusion can be summarised as: "Don't live to work but work to live!". Participants spontaneously rearranged the chairs to form a big circle so as to facilitate discussion and be able to see everyone, an arrangement that was then reproduced for each round table.

Wednesday morning's talks targeted the cosmic distance scale and stellar populations. Again we saw great scientific results thanks to the synergy between different organisations and facilities. The afternoon was dedicated to a round-table discussion on Diversity-Equity-Inclusion, chaired by Belén Alcalde and with several panellists (see Alcalde Pampliega, Hibon & de Gregorio Monsalvo on p. 46 of this issue for more details). Everyone showed genuine concern for this topic and agreed it should not be a side-issue. We must educate ourselves in this area to ensure progress.

Thursday's presentations concentrated on extragalactic astronomy: the high-redshift Universe, active galactic nuclei, and large-scale structure. We learned about impressive results from programmes at many different observatories, including some Large Programmes, and the power of combining ALMA and other observations (ground- or space-based) for making progress on important problems in the area galaxy formation and evolution. Through these efforts, the high-redshift Universe is revealing itself more and more.

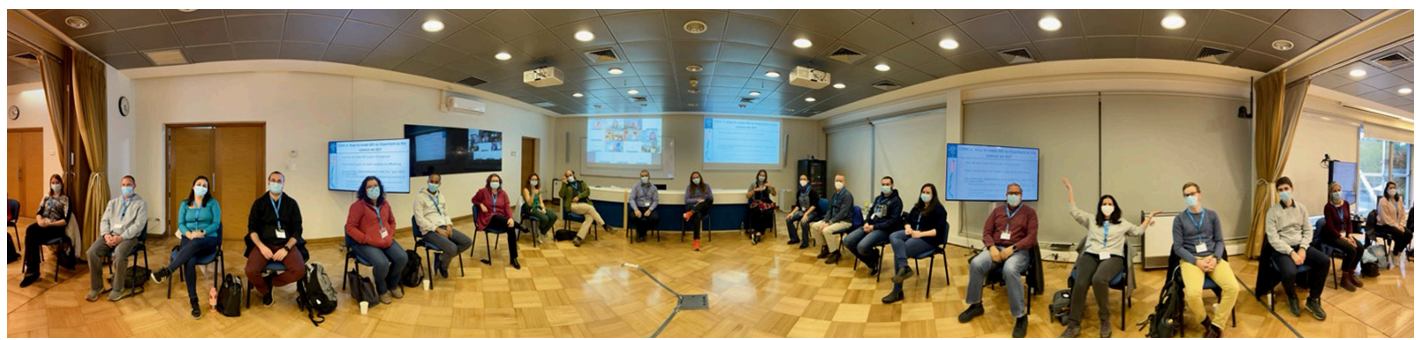

Figure 1. Round table photo. This room configuration optimised the discussions.

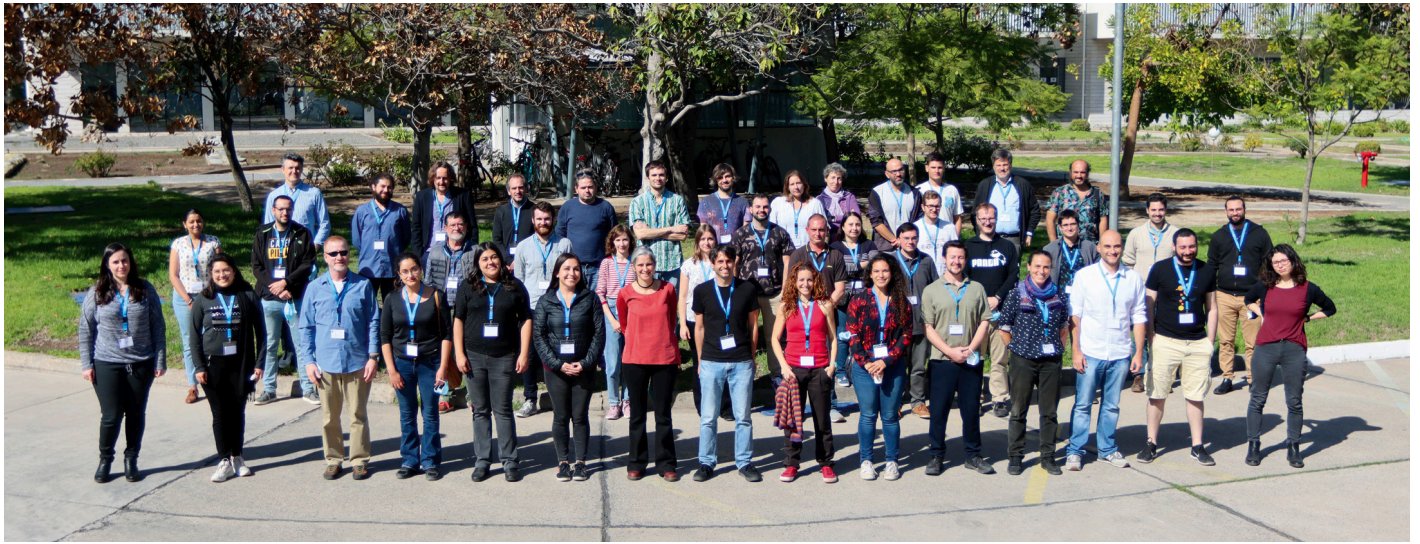

Figure 2. Conference photo.

On Friday morning the talks were centred on technical developments and some issues that we all face at observatories, including the latest astronomical technologies developed in Chile (Polymer Reinforced Carbon Fiber mirrors, adaptive optics) and the impact of low-Earth-orbit satellites on nighttime observations. The afternoon was dedicated to a round-table discussion, chaired by Franz Bauer, with the directors of the Chile-based observatories: Andreas Kaufer (ESO-LPO Director), Leopoldo Infante (LCO Director), Elizabeth Humphreys (ALMA Head of Science Operations), Robert Blum (Director for Operations at Vera C. Rubin Observatory), and the chair of the Chilean Telescope Allocation Committee, Patricio Rojo. Discussions were focused on the road ahead. Panellists and participants engaged in lively discussion of the new projects, observatory sustainability, remote working and the several opportunities to link the different observatories and the Chilean institutes. We all agreed on the efforts and actions that needed to be undertaken to involve the Chilean community and to retain minorities.

Although the weather during the week was highly variable, the quality of the food at each coffee break and lunch, provided by new catering companies that are all managed by Chilean women, spoiled us. The feedback from speakers, panellists, and participants was extremely positive and most of the audience asked for a repeat of the forum. This will allow continued discussion of the progress made, not only from the scientific perspective, but especially on the topics from the different round tables. Everyone left the meeting with a big smile.

## Demographics

As with many workshops, the Science Organising Committee sought fair representation from the community. To this end, we voted on 16 invited speakers, using the sole criterion of who would give the best review of each topic. The end result was a 50:50 ratio of male to female invited speakers, with four PhD students delivering invited talks. The Diversity panel had a 50:50 ratio of male to female panellists, The Life Balance panel had a 20:80 ratio of male to female panellists, and The Road Ahead panel an 80:20 ratio of male to female panellists. This last does reflect the actual ratio of males to females in observatory senior management. Attendees came from all Chile-based astronomical observatories and institutions with the following percentages:
– 58% from observatories (ESO, ALMA, NOIRLab, LCO, Carnegie, GMT);
– 42% from Chilean astronomical universities/institutions.

Of the abstract submissions, 35% were from women, which matched the 35 % of talk allocations to women. The talk selection was made blind (the SOC chair removed first names and identifying information about the authors), so we conclude that the method likely worked to address gender biases. We also had a decent level of participation from young researchers, within the following breakdown according to seniority: ~ 14% students, ~ 18% postdoctoral researchers, 15% engineers and operators, and 48% tenure-track or tenured astronomers and 5% outreach or HR professionals. Each of these groups was well represented in the talks, including the invited talks.

The workshop had a high level of participation, with approximately 110 participants. We attribute this to both the compelling nature of the subject matter, which draws researchers at all career stages, and to the generous support that ensured attendance was completely free.

### Acknowledgements

We would like to acknowledge and sincerely thank Christopher Martin and the Kavli Foundation who provided the funding for this forum and without whom this meeting would not have been possible. We also would like to acknowledge the forum proposal Co-Is, the SOC, the LOC and the ESO logistics team. We also would like to thank the ESO Chile Office for Science and Logistics for their support with the organisation of this forum.